\documentclass[aps,prd,twocolumn,superscriptaddress,floatfix,showpacs,showkeys,preprintnumbers]{revtex4}

\usepackage[utf8]{inputenc}
\usepackage[T1]{fontenc}
\usepackage{slashed}
\usepackage{times}

\usepackage{amssymb,amsfonts,amsmath,amsthm}
\usepackage{dsfont,bbm}
\usepackage{dcolumn}

\usepackage{epsf}

\usepackage{graphicx}
\usepackage[caption=false]{subfig}

\usepackage{dsfont}

\usepackage{epstopdf}

\usepackage{slashed}

\usepackage[active]{srcltx}

\usepackage[usenames]{color} 

\begin{document}


\title{Axial form factor of the nucleon at large momentum transfers}

\author{I.V.~Anikin}
\affiliation{Institut f\"ur Theoretische Physik, Universit\"at
   Regensburg,D-93040 Regensburg, Germany}
\affiliation{Bogoliubov Laboratory of Theoretical Physics, JINR, 141980 Dubna, Russia}
\author{V.M.~Braun}
\affiliation{Institut f\"ur Theoretische Physik, Universit\"at
   Regensburg,D-93040 Regensburg, Germany}
\author{N.~Offen}
\affiliation{Institut f\"ur Theoretische Physik, Universit\"at
   Regensburg,D-93040 Regensburg, Germany}

\date{\today}

\begin{abstract}
  \vspace*{0.3cm}
\noindent
Motivated by the emerging possibilities to study threshold pion electroproduction at large 
momentum transfers at Jefferson Laboratory following the 12 GeV upgrade, we provide a 
short theory summary and an estimate of the nucleon axial form factor for large virtualities in 
the $Q^2 = 1-10~\text{GeV}^2$ range using next-to-leading order light-cone sum rules.  
 \end{abstract}



\maketitle

\date{\today}

\section{Introduction}
\setcounter{equation}{0}

The structure of the nucleon probed by the axial current is described by two form factors
that are conventionally defined as
\begin{eqnarray}
\lefteqn{
 \langle N(p') |j_{\mu5}^a|N(p)\rangle =}
\nonumber\\&=& \bar u(p')\Big[\gamma_\mu G_A(Q^2) + 
\frac{(p'-p)_\mu}{2m} G_P(Q^2)\Big]\frac{\tau^a}{2}u(p)\,,
\end{eqnarray}  
where $j_{\mu5}^a = \bar q \gamma_\mu\gamma_5 \frac{\tau^a}{2} q$ is the $SU(2)$-flavor 
isovector axial current, $\tau^a$ are the usual Pauli matrices, $Q^2=-(p'-p)^2$ the invariant momentum transfer squared, 
and $m=(m_p+m_n)/2$ the nucleon mass.

The axial form factor $G_A(Q^2)$ can be determined either from 
quasi-elastic neutrino scattering or from threshold pion electroproduction (with the help of
current algebra and chiral perturbation theory). The experimental results and theory methods used 
for their extraction at low-to-moderate momentum transfers can be found in the excellent 
review~\cite{Bernard:2001rs} and need not to be repeated here. All existing data at 
$Q^2 \lesssim 1$~GeV$^2$ can be described by the dipole formula
\begin{equation}
 G_A(Q^2) = \frac{g_A}{(1+Q^2/M^2_A)^2}\,,
\label{dipole}
\end{equation} 
where $g_A=1.2673(35)$ is the axial-vector coupling constant and the parameter $M_A$, 
the so-called axial mass, is determined to be~\cite{Bernard:2001rs}
\begin{eqnarray}
  M_A &=& 1.026(21)~\text{GeV}\qquad(\text{neutrino~scattering})\,,
\nonumber\\
  M_A &=& 1.069(16)~\text{GeV}\qquad(\text{electroproduction})\,.
\label{MA}
\end{eqnarray}
Taken literally the difference between the two axial mass determinations by these two techniques translates to
a difference of about 5\% for the nucleon axial radius. Resolution of this discrepancy is 
discussed in detail in~\cite{Bernard:2001rs,Bernard:1995dp}.
The induced pseudoscalar form factor $G_P(Q^2)$ is believed to be understood in terms of the pion pole
dominance up to small corrections~\cite{Bernard:2001rs}.  

It has been pointed out that the dipole ansatz can be too restrictive, and hence the errors
underestimated. This affects potentially both small- and large-$Q^2$ extrapolations.
The most recent neutrino data analysis in a broader $Q^2$ region using a more flexible $z$-parametrization is 
presented in~\cite{Meyer:2016oeg}.

The motivation for our work comes from the emerging possibilities to study threshold pion 
electroproduction at large momentum transfers at Jefferson Laboratory following the 12 GeV upgrade.
Such data already exist for the $Q^2\sim 2-4$~GeV$^2$ range~\cite{Park:2012yf} but up to now remain largely unnoticed. 

We remind that the extraction of the axial form factor from pion electroproduction goes back to the classical 
low-energy theorem of Nambu, Luri\'e and Shrauner for the electric dipole amplitude $E_{0+}$ at 
threshold~\cite{Nambu:1997wa,Nambu:1997wb}. In the strict chiral limit $m_\pi=0$ this theorem is 
valid for arbitrary momentum transfer $Q^2$. However, the finite pion mass corrections $m_\pi/m \simeq 1/7$  are tricky.
They can be calculated reliably at small $Q^2 \sim (100~\text{MeV})^2$ using the low-energy effective theory~\cite{Bernard:2001rs,Bernard:1995dp},
but are model-dependent beyond this range. Several models for such corrections were developed to connect low energy theorem 
and data. 

The extension to the large $Q^2$ region is not straightforward because the theoretical limits $Q^2\to\infty$ and $m_\pi\to 0$ do not 
commute, in general. In physics terms the problem is that at large momentum transfers the emitted pion cannot be soft to both the 
initial and the final nucleon simultaneously. As a result classical low-energy theorems are expected to break down at 
$Q^2 \sim m^3/m_\pi$~\cite{Pobylitsa:2001cz}: the initial-state pion radiation occurs at time scales of
order $1/m$ rather than $1/m_\pi$ necessitating to add contributions of hadronic intermediate states other
than the nucleon. The analysis in~\cite{Braun:2006td,Braun:2007pz} suggests that such corrections to the transverse $E^{(-)}_{0+}$ amplitude
remain small, of the order of 20\%, in the the whole region  $Q^2\sim 1-10$~GeV$^2$ that is interesting in view of the 
forthcoming JLAB12 experimental program whereas the longitudinal $L_{0+}$ amplitude appears to be much stronger affected.     
From this evidence the worst case scenario seems to be that finite pion mass corrections to the nucleon axial form factor extractions from the 
threshold pion productions using low-energy theorems can reach 30\%. Such corrections can be, however, estimated within models 
so that we expect that with some more theory input the remaining uncertainty can be brought below 10-15\%. 
Whereas this accuracy may not seem attractive for the 
low $Q^2$ range, it is certainly interesting  for studies in the few GeV region and will be challenging to match 
with similar experimental precision.      
In our opinion such measurements would be very interesting and the task of this note is to provide one with the corresponding QCD expectations.   
To this end we present a calculation of the nucleon axial form factor for photon virtualities in the $Q^2 = 1-10~\text{GeV}^2$ range using 
next-to-leading order (NLO) light-cone sum rules. 

\section{Light-cone sum rules}
\setcounter{equation}{0}  

It is generally accepted that hadron form factors in the formal $Q^2 \to \infty$
limit are dominated by hard gluon exchanges between the valence quarks at small transverse
separations. 
However, there is overwhelming evidence that the hard rescattering  regime is not achieved for realistic 
momentum transfers accessible at modern accelerators and the so-called ``soft'' or Feynman-type
contributions play the dominant role. 
Soft contributions can be estimated using the light-cone sum rule (LCSR) technique 
that is based on the light-cone operator product expansion of suitable correlation functions 
combined with dispersion relations and quark-hadron 
duality. This technique is attractive because it can be applied to all elastic and transition form factors and 
involves the same universal nonperturbative functions that enter the pQCD calculation; there
is no double counting and (almost) no new parameters. 

The LCSRs for the electromagnetic nucleon form factors have been derived in~\cite{Braun:2001tj,Braun:2006hz} to the 
leading order (LO) and recently in~\cite{Anikin:2013aka} to the NLO in the QCD coupling. For the axial form factor
only the LO results are available~\cite{Braun:2006hz,Aliev:2007qu}. It turns out, however, that the NLO LCSRs 
for the axial form factor do not require a new calculation and can be obtained using the expressions presented 
in~\cite{Anikin:2013aka} with minor modifications.    

The starting point is the correlation function 
\begin{equation}
 T_{\mu5}(p,q) = i\int \!d^4x\, e^{iqx} \langle 0| T\{\eta(0)j_{\mu5}(x)\}|P(p)\rangle\,,
\end{equation}
where $|P(p)\rangle$ is the proton state with momentum $p_\mu$, $p^2=m^2$, and $\eta(x)$ is a 
suitable local operator with proton quantum numbers (Ioffe current).
The corresponding coupling is  $\langle 0|\eta(0)|P(p)\rangle  = \lambda_1\, m\, u(p)$.
For technical reasons it is more convenient to consider the neutral axial vector current
\begin{equation}
 j_{\mu 5} = \frac12 \big[ \bar u \gamma_\mu\gamma_5 u - \bar d \gamma_\mu\gamma_5 d \big]. 
\end{equation}

Following~\cite{Braun:2001tj} we consider the ``plus'' projection of the correlation function
in the Lorentz and spinor indices which can be parametrized by 
two invariant functions:
\begin{equation}
 \Lambda_+ T_{+5} = p_+ \big[ m \, \mathcal{A}_5 (Q^2, p'^2) + \slashed{q}_\perp \mathcal{B}_5 (Q^2, p'^2)\big]\gamma_5 u_+(p)\,,
\end{equation}  
where $p'=p+q$. The invariant functions can be calculated for large Euclidean momenta $Q^2, -p'^2 \ll \Lambda_{\rm QCD}^2$ 
using the light-cone OPE. The results can be written in the form of a dispersion integral
\begin{equation}
 \mathcal{A}_5^{\rm QCD} (Q^2, p'^2) = \frac1{\pi}\int_0^\infty \frac{ds}{s-p'^2}\,\text{Im}\,\mathcal{A}_5^{\rm QCD}(Q^2, s) +\ldots
\label{A:QCD}
\end{equation}  
where $\text{Im}\, \mathcal{A}^{\rm QCD} (Q^2, s)$ is given by the convolution of perturbatively calculable coefficient 
functions $C_5^{\mathcal{F}}$ and the matrix elements of three-quark operators at light-like separations, $\mathcal{F}(x,\mu_F)$,
dubbed distribution amplitudes (DAs),
\begin{equation}
 \text{Im}\, \mathcal{A}_5^{\rm QCD} = \sum_{\mathcal{F}} C_5^{\mathcal{F}}(x,Q^2,s,\mu_F,\alpha_s(\mu_F))\otimes \mathcal{F}(x,\mu_F)\,.  
\end{equation}
The sum goes over all existing DAs of increasing twist, $x=\{x_1,x_2,x_3\}$ stands for the quark momentum fractions 
and $\mu_F$ is the factorization scale. 

Leading-order (LO) expressions are available from~\cite{Braun:2006hz}, see Eq.~(A.7).
For consistency with our NLO calculation we expand all kinematic factors in the LO results in
powers of $m^2/Q^2$ and neglect terms $\mathcal{O} (m^4/Q^4)$. This truncation is also consistent with taking
into account contributions of twist-three, -four, -five (and, partially, twist-six) in the OPE. 
The NLO expressions for $\mathcal{A}_5$ can be obtained from the results in~\cite{Anikin:2013aka} (see Appendix~E) with
the following replacements:
\begin{itemize}
\item{} For the $d$-quark contribution, replace $e_d \to 1/2$.
\item{} For the $u$-quark contribution,  replace $e_u \to 1/2$ and interchange symmetric and antisymmetric parts of the 
 DAs: $\mathcal{V}_1 \leftrightarrow - \mathcal{A}_1$, $\mathcal{V}_2 \leftrightarrow  \mathcal{A}_2$,
   $\mathcal{V}_3 \leftrightarrow \mathcal{A}_3$.
\end{itemize}  
The sum rules are constructed by matching the QCD representation (\ref{A:QCD}) to the dispersion representation 
in terms of hadronic states
\begin{align}
  \mathcal{A}_5^{\rm QCD} (Q^2, p'^2) &= \frac{2\lambda_1 G_A(Q^2)}{m^2-p'^2}
\\& \quad
+ \frac1{\pi}\int_{s_0}^\infty \frac{ds}{s-p'^2}\,\text{Im}\,\mathcal{A}_5^{\rm QCD}(Q^2, s) +\ldots
\nonumber
\end{align}
where it is assumed that contributions of nucleon resonances and scattering states are effectively
taken into account by the QCD expression above a certain threshold $s_0\simeq (1.5\,\text{GeV})^2$
(interval of duality). Applying the Borel transformation $p'^2\to M^2$ to get rid of the subtraction 
constants and suppress higher-mass contributions, one obtains the LCSR~\cite{Braun:2001tj,Braun:2006hz}
\begin{align}
 2\lambda_1 G_A(Q^2) &= \frac1{\pi}\int_{0}^{s_0} \!ds\,e^{(m^2-s)/M^2}\,\text{Im}\,\mathcal{A}_5^{\rm QCD}(Q^2, s)
\label{GA:LCSR}
\end{align}  
that we analyze in what follows~\cite{remark1}. 
 
\section{Results}\label{sec:Results}

\begin{figure*}[ht]
  \begin{center}
\includegraphics[width=0.47\linewidth]{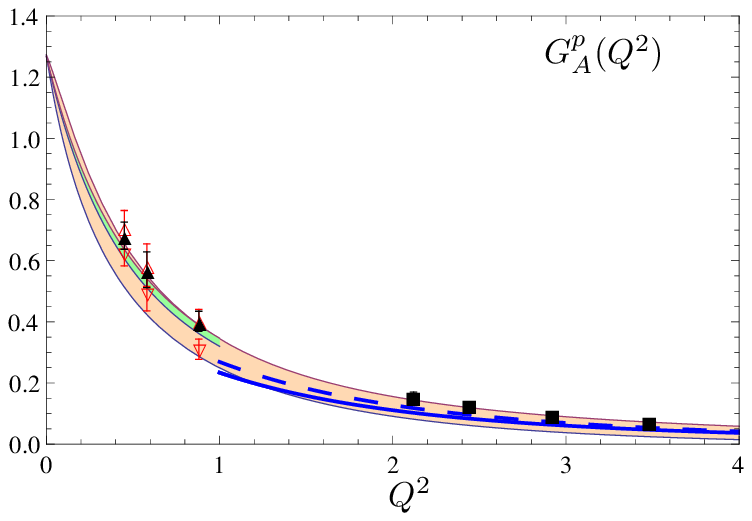}
\qquad
\includegraphics[width=0.47\linewidth]{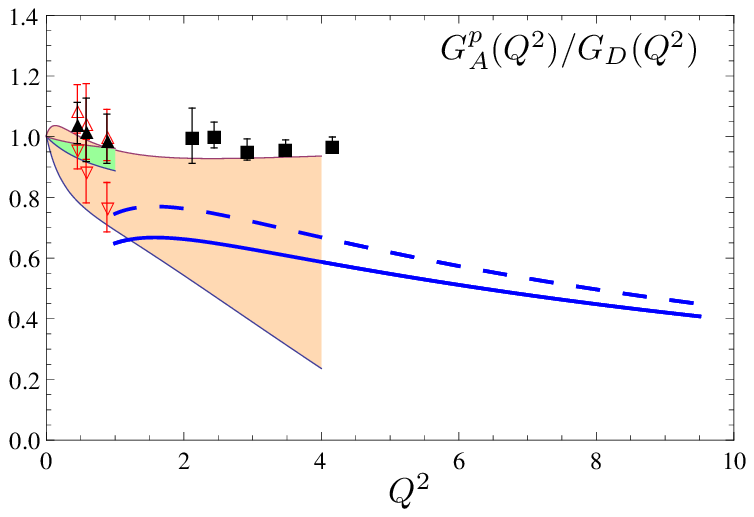}
   \end{center}
\caption{Axial form factor of the nucleon from LCSRs compared to the experimental data.
Parameters of the nucleon DAs correspond to the sets ABO1 and ABO2 in Table~I of Ref.~\cite{Anikin:2013aka} 
for the solid and dashed curves, respectively. Borel parameter $M^2=1.5$~GeV$^2$ for ABO1 and $M^2=2$~GeV$^2$ for ABO2.
The dipole- and z-parametrizations of the neutrino scattering data are shown by the narrow (green) and broad (orange) shaded regions, respectively.
The data points are from the pion electroproduction experiments, Refs.~\cite{DelGuerra:1976uj,Park:2012yf}. For more details see text. 
}
\label{fig:ABOaxial}
\end{figure*}

The results are shown in Fig.~\ref{fig:ABOaxial} where on the left panel we plot $G_A(Q^2)$ in absolute normalization
and on the right panel the ratio of  $G_A(Q^2)$ to the dipole formula (\ref{dipole}) with the axial mass $M_A = 1.069$~GeV corresponding to
the average value (\ref{MA}) from the pion electroproduction measurements~\cite{Bernard:2001rs}.   

%


The LCSR for the axial form factor in Eq.~(\ref{GA:LCSR}) does not contain free parameters.
The results are shown for two realistic models of the leading- and higher-twist 
nucleon DAs,  ABO1 (solid curves) and ABO2 (dashed curves), defined in in Table~I of Ref.~\cite{Anikin:2013aka}.  
These models have been obtained by combining the available lattice QCD constraints~\cite{Braun:2008ur,Braun:2014wpa} 
with the fit to the electromagnetic proton form factors, $F_1(Q^2)$ and $F_2(Q^2)$, see Fig.~3 in~\cite{Anikin:2013aka}. 
The NLO corrections that are the subject of this work are large and positive (up to 40\%) at $Q^2=1-2$~GeV$^2$ but decrease 
(to below 15\%) at larger momentum transfers and change sign at $Q^2\sim 6$~GeV$^2$ for the both DA models.     
In addition to the uncertainty in the nonperturbative input, there exist also intrinsic uncertainties of the LCSR approach 
itself (factorization scale and Borel parameter dependence, higher-order and higher-twist corrections etc.) 
that we estimate to be 10-15\%. We, therefore, expect the overall accuracy 
of our predictions for the axial form factor in the optimal for this technique range $Q^2\sim 3-10$~GeV$^2$ to be of the order of $20-25\%$.  
We show the results starting at $Q^2 > 1$~GeV$^2$. However, the experience with the LCSRs for $B$-decay and nucleon 
electromagnetic form factors indicates that momentum transfers in the 1-2~GeV$^2$ range are still too low 
for a fully quantitative treatment in this approach.

A compilation of the low-$Q^2$ measurements can be found in~\cite{Bernard:2001rs} (see also \cite{Meyer:2016oeg}).  
For the neutrino scattering, in order not to overload the plot we show  the standard dipole parametrization with the axial mass 
$M_A = 1.026(21)$ GeV by the narrow shaded area, and, in addition, by a broader shaded area extending to $Q^2=4$~GeV$^2$, 
the one-sigma envelope from the recent analysis using a more general z-parametrization that also includes newer deuterium 
data~\cite{Meyer:2016oeg}.
For the same reason we do not show ``old'' 
electroproduction data except for~\cite{DelGuerra:1976uj} in the range $Q^2 = 0.45-0.88$ GeV$^2$.
The three shown sets of data points correspond to the form factor extraction using the strict soft-pion limit (filled triangles) 
and two models for the hard pion corrections (open triangles). 
The recent CLAS data~\cite{Park:2012yf} are shown by filled squares. These results were obtained by employing the low-energy theorem 
in the chiral limit and extracting the $E_{0+}$ multipole from the fit to the total cross section $\gamma^*p \to\pi^+n $ at the 
energy $W=1.11$~GeV, closest to the threshold. 
Our predictions for the large $Q^2$ region match the existing neutrino scattering data at smaller momentum transfers~\cite{Meyer:2016oeg} very well,  
and are about 20-30\% below the CLAS extraction from pion electroproduction in the soft pion limit~\cite{Park:2012yf}. Since the corrections
to the  soft pion limit are expected to be negative~\cite{DelGuerra:1976uj,Braun:2006td,Braun:2007pz} and can well be in the 20\% range,
there is no contradiction. A more detailed analysis of such corrections within realistic models would be very welcome. 

To summarize, we argue that studies of pion electroproduction at threshold $\gamma^*p\to \pi^+n$ at large photon virtualities accessible at the 
Jefferson Laboratory following the 12 GeV upgrade supplemented by the measurements of the neutron magnetic form factor in the same $Q^2$ range
provide one with a viable method to determine the axial proton form factors with the theoretical accuracy that is currently limited to 
$20-30\%$ but, very likely, can be improved in the future. These results can be confronted with QCD predictions based on LCSRs and, potentially,
lattice QCD (e.g.~\cite{Alexandrou:2010hf,Junnarkar:2014jxa}) and Dyson-Schwinger equations~\cite{Eichmann:2011pv} 
although the extension to the large-$Q^2$ region in both approaches can be challenging.        
A combination of lattice calculations with models can offer additional insights, e.g.~\cite{Ramalho:2015jem}. 

\vspace*{2mm}

\section*{Acknowledgments}
The authors are grateful to Kijun Park for the tables of the experimental data reported in \cite{Park:2012yf}.
Special thanks are due to Richard Hill and Aaron Meyer for bringing Ref.~\cite{Meyer:2016oeg} to our attention
and correspondence.
The work by I.A. was partially supported by the Heisenberg--Landau Program of the German Research Foundation (DFG).






\end{document}